\documentclass[preprint,showpacs,preprintnumbers,amsmath,amssymb]{revtex4}
\usepackage{graphicx}
\usepackage{dcolumn}
\usepackage{bm}
\usepackage{hyperref}
\newcommand{\hodge}[1]{\,*#1}
\newcommand{\lie}{{\cal L}}
\begin{document}
\preprint{}
\title{Charges and the boundary in Chern Simons gravity}
\author{Rodrigo Aros}
\affiliation{Departamento de Ciencias F\'{\i}sicas, Universidad Andr\'es Bello, Av. Republica 252,
Santiago,Chile}

\date{\today}
\begin{abstract}
In this work a new set of boundary conditions for Chern Simons gravity that lead to a fully gauge
invariant action is analyzed. This particular form of the action reproduces the standard results
of black hole thermodynamics and determines that the algebra of charges of diffeomorphisms at the
horizon be the Virasoro algebra.
\end{abstract}
\maketitle

\section{Introduction}
For all known classical and quantum theories boundary conditions have a prominent role, however,
for gravity they are harder to analyze since gravity must determine the space itself and so the
boundary. As a rule to overcome this cumbersome problem the geometry of the asymptotical regions
is imposed \textit{a priori}. Furthermore, it is usually imposed that the space be asymptotically
(locally) maximally symmetric, \textit{i.e.}, either flat or (Anti) de Sitter (\textbf{(A)dS}).

The astrophysical evidence in four dimensions supports a positive cosmological constant, however
in higher dimensions there are several theoretical reasons to consider a negative cosmological
constant. Besides the famous AdS/CFT correspondence \cite{Maldacena:2000mw} one can mention for
instance Ref.\cite{Friedman:1993ty} where scalar fields, for a range of masses, have two different
\textit{quantization }permitted, or the algebraic relations, otherwise differential, in the
holographic renormalization \cite{Skenderis:2002wp} prescription. In this work, and for
simplicity, only the AdS group will be addressed.

The presence of a negative cosmological constant, however, introduces some technicalities in the
boundary conditions which if are ignored leads to inconsistences, for instance, to  divergent
charges. To address that several different regularization procedures have been proposed (see for
instance \cite{Hawking:1996fd,Balasubramanian:1999re}), where each procedure is somehow connected
with a different boundary condition. In even dimensions for a negative cosmological constant the
ALAdS boundary condition was introduced in \cite{Aros:1999id}. Under this condition the action is
tailored to be -on shell- stationary for arbitrary variations of the fields on the asymptotical
spatial region which is locally anti de Sitter. In odd dimensions a similar procedure has been
more elusive, although some important results has been obtained \cite{Mora:2004kb, Mora:2004rx}.
The intension of this work to shed some more light on this case.

Standard gravity stands apart from the rest of the interactions in at least one fundamental point,
gravity is not a gauge theory. In three dimensions, however, the Einstein theory, properly
reformulated in first order, can be interpreted as a Chern Simons theory \cite{Achucarro:1987vz}
which is a genuine gauge theory.

The Chern Simons (\textbf{CS}) theories are gauge theories, different from Yang Mills ones, whose
Lagrangians are Euler-Chern-Simons densities for the corresponding gauge group, thus they exist
only in odd dimensions. In fact in any $d>3$ odd dimension there exists a CS theory of gravity
\footnote{The CS Lagrangian belongs to a larger family of theories of gravity, the Lovelock
gravities \cite{Lovelock:1971yv}, each one including higher power of the Riemann tensor but with
second order equations of motion for the metric.}, which differs from the Einstein one.
Furthermore, there are Chern Simons SUGRA theories in any odd dimension \cite{Troncoso:1998ng}.

Unlike the Yang Mills Lagrangian an Euler Chern Simons density is a quasi-invariant form ,
\textit{i.e.} it changes in a boundary term under gauge transformations. Precisely this lack of
full gauge invariance is remedied in this work. To do that one can recall that originally the
Chern Simons action arises in the context of anomalies as the reduction of the gauge invariant
form, known as transgression \cite{AnomaliesBook,ThesisPablomora,Mora:2006ka,Mora:2006tq},
\begin{equation}\label{Trans}
I_{2n+1} = (n+1) \int_{\mathcal{M}} \int^{1}_{0}dt \left\langle(A_{1}-A_{0})\wedge
\underbrace{F_{t}\wedge\ldots\wedge F_{t}}_{n}\right\rangle,
\end{equation}
where $A_{1}$ and $A_{0}$ are two (1-form) connections in the same fiber. $F_{t}=dA_{t} +
A_{t}\wedge A_{t}$ with $A_{t} = t A_{1} + (1-t) A_{0}$. $\langle \rangle$ is the trace in the
group. From now on, the differential forms language is assumed, thus $\wedge$ product will be
implicit.

The Euler Chern Simons density for $A_{1}$ is obtained from Eq.(\ref{Trans}) by setting $A_{0}=0$,
or viceversa. Particularly the transgression form (\ref{Trans}) can be reduced to the Euler Chern
Simons form in any charts of the fiber bundle but not globally unless the fiber bundle be trivial
\cite{AnomaliesBook}.

One can show the manifest gauge invariance of Eq.(\ref{Trans}) by noting that under a gauge
transformation with group element $g$,
\begin{equation}\label{gaugetransformation}
A_{1}-A_{0} \mapsto g^{-1}(A_{1}-A_{0})g  \textrm{ and } F_{t} \mapsto g^{-1} F_{t} g,
\end{equation}
which together with the invariance of the trace, $\langle \rangle$, under cyclic rotations confirm
the manifest invariance of Eq.(\ref{Trans}) under gauge transformations.

The final intension of this work is to study some of consequences of a fully gauge invariant Chern
Simons theory, in particular in gravity. In the standard Chern Simons theory is well established
that because the action is a quasi invariant form part of the group elements $g$ in the gauge
transformations becomes dynamical at the boundary introducing new degrees of freedom  (See for
instance \cite{Witten:1988hf,Banados:1996tn}) into the theory.  Conversely, for an
\textit{improved}, and fully gauge invariant action, no new degree of freedom can arise at the
boundary. Therefore it is very interesting to address this fundamental difference. Almost every
result, beyond black hole solutions, in Chern Simons gravity -entropy, central charges, etc.- is
usually connected with this lack of full gauge invariance. As shown in the next sections, the
truly gauge invariance of the action ($\ref{Trans}$) has some significant consequences. Although
it is not investigate in this work it is very important to note that since the action
($\ref{Trans}$) is by construction manifestly invariant under gauge transformations then much of
the rich structure associated with the breaking or preservation of gauge symmetries may be
different in this case. Unfortunately the dynamics of Chern Simons theories is just being starting
to be understood \cite{Banados:1996yj,Miskovic:2005di}.

The first part of this work is the redefinition, or reinterpretation, of the action (\ref{Trans})
as a standard Chern Simons action with particular set of boundary conditions through the election
of a particular $A_{0}$ field without loosing the manifest gauge invariance of the theory. It is
worth to stress that this redefinition solves the ambiguity of having two dynamical fields, which
otherwise might even forbid a Hamiltonian version of the theory.

As mention before in terms of Chern Simons gauge theory, unlike Yang Mills ones, one can construct
a theory of gravity. Because of that the rest of the work is devote to the analysis of two other
different aspects of Eq.(\ref{Trans}) as representing a theory of gravity are discussed. The first
aspect is the thermodynamics of its black holes which in this work is proven to reproduce the
standard black hole thermodynamics obtained in Refs.\cite{Banados:1994ur,Aros:2000ij}. Also in
this part is proven that Hamiltonian and Noether charges agree because of the boundary conditions.

The second aspect, which is connected with the previous one but one can consider it more
fundamental, is the structure of charges, i.e., the algebras its charges realize. In this case is
proven that the standard algebras for diffeomorphisms and gauge are recovered, however without
central extensions unlike Refs.\cite{Brown:1986nw,Banados:1996tn}.

In this work space considered is manifold $\mathcal{M}$. It has two boundaries, an outer region
with corresponds to the asymptotical spatial infinity, and since black holes will be discussed, an
internal boundary which represents the horizon. $\mathcal{M}$ can be pictured as
$\mathcal{M}=\mathbb{R}\times \Sigma$ where $\Sigma$ corresponds to a $2n$-dimensional spacelike
hypersurface and $\mathbb{R}$ stands for the time direction. The boundary involving the horizon
will be denoted as $\mathbb{R}\times
\partial\Sigma_{H}$ thus $\partial\mathcal{M}\mathbb{=R}\times
\partial\Sigma_{\infty} \cup \mathbb{R}\times
\partial\Sigma_{H}$.

\section{Background fields and boundary terms}

One can readily check that the variation of Eq.(\ref{Trans}) yields
\begin{equation}\label{variationoftrans}
\delta I_{2n+1} = \int_{\mathcal{M}} \langle F^{n}_{1} \delta A_{1}\rangle + \langle F^{n}_{0}
\delta
 A_{0}\rangle + d\Theta
\end{equation}
where the boundary term reads
\begin{equation}\label{Boundarytem}
    \Theta = -n(n+1)\int^{1}_{0}dt \, \langle (A_{1}-A_{0}) F_{t}^{n-1} \delta A_{t} \rangle.
\end{equation}
Note that equation of motion for $A_{1}$ or $A_{0}$ are the  equation of motion for the standard
Chern Simons action.

The interpretation of Eq.(\ref{Trans}) requires a discussion (See
\cite{ThesisPablomora,Mora:2006ka,Mora:2000me,Mora:2000ts}) since the presence of two different
fields, $A_{0}$ and $A_{1}$, may be confusing. This is solved if one consider $A_{0}$ as a
background field and $A_{1}$ as the dynamical one, or viceversa. In this work $A_{0}$ will be the
background field. To clarify this idea one must note that Eq.(\ref{Trans}) can be rewritten as
\begin{equation}\label{TransRewritten}
  I_{2n+1} = I_{CS}(A_{1}) - I_{CS}(A_{0}) + \int_{\partial \mathcal{M}} B(A_{1},A_{0},F_{t}),
\end{equation}
where $I_{CS}$ stands for the standard CS action and $B$ for a boundary term.

To impose $A_{0}$ as a background naturally leads to fix $A_{0}$, module gauge transformations.
However to avoid ambiguities, in particular if one would like to construct a theory of gravity
from Eq.(\ref{Trans}), one needs that $A_{0}$ has no influence at the bulk but only at the
boundary of the manifold. To realize that one could set $A_{0}=0$ in the bulk and $A_{0}\neq 0$
only at the boundary as done in Ref.\cite{Mora:2004kb,Mora:2006ka} because of in that case
$I_{CS}(A_{0}=0)=0$. Unfortunately that configuration is a little \textit{tricky} because
$A_{0}=0$ is a not a gauge invariant \textit{state}, since any gauge transformation generates
$A_{0}=0 \mapsto A_{0}=g^{-1}dg$. Furthermore in the next paragraph is shown that imposing
$A_{0}=0$ at the bulk probably disregards some of the structure of the Chern Simons theory here
proposed. Because of that, and to gain generality, instead in this work $A_{0}=g^{-1}dg$ is chosen
everywhere (bulk and boundary). As shown next this does not introduce contributions at the bulk
classically. Note that the formal expression $A_{0}= \hat{g}^{-1} d\hat{g}$ is preserved under
gauge transformations.

The ambiguity between $A_{0}=0 \Longleftrightarrow A_{0}=g^{-1}dg$ fortunately can be solved by
the Chern Simons action itself. The Euler Chern Simons density $I_{CS}(A_{0})$ can be rewritten as
a polynomial of $A_{0}$ and $F_{0}$
\[
I_{CS}(A_{0}) = \int_{\mathcal{M}} \sum_{p=0}^{n} \alpha_{p} \left\langle A_{0}^{2(n-p)+1}
F^{p}_{0}\right\rangle,
\]
where $\alpha_{p}$ are fixed constants. However for $A_{0}=g^{-1}dg$, a  \textit{flat} connection
(locally $F_{0}=0$), the CS action becomes
\begin{equation}\label{WZWHigher}
  I_{CS}(g^{-1}dg) = \alpha_{0} \int_{\mathcal{M}} \left\langle (g^{-1}dg)^{2n+1}\right\rangle.
\end{equation}
Eq.(\ref{WZWHigher}) is closed and its variation is a boundary term, thus its addition does not
introduce equations of motion in the bulk as expected. Note also that Eq.(\ref{WZWHigher}) is a
generalization of the 2+1 Wess-Zumino term.

These definitions imply that the action (\ref{Trans}), provided $A_{0}=g^{-1}dg$, can be
interpreted as standard Chern Simons action for $A_{1}$ with a boundary term $B(A_{1},g)$ and a
WZW$_{2n+1}(g)$ term that make it manifestly gauge invariant.

\section{AdS Gravity}

Although the structure of (\ref{Trans}) as a gauge theory is interesting by it self, from now on
this work is devoted to the analysis of (\ref{Trans}) as a theory of gravity.

The Euler Chern Simons form, and so Eq. (\ref{Trans}), can represent a theory of gravity if the
underlying gauge group is either Poincar\'{e}, AdS or dS group in the corresponding dimension. The
different groups are connected with a vanishing, negative or positive cosmological constant
respectively \cite{Zanelli:2002qm}.

Basically to construct the theory of gravity for the AdS group either $A_{0}$ or $A_{1}$ have the
generic form
\begin{equation}\label{A}
    A = \frac{1}{2} \omega^{ab} J_{ab} + \frac{e^a}{l} P_{a},
\end{equation}
where $e^{a}$ is a vielbein and $\omega^{ab}$ a connection for the local Lorentz group in the
(co)tangent space. In this theory the connection and the vielbein are independents fields. $P_{a}$
and $J_{ab}$ are the generator of the AdS group ( see appendix (\ref{ads})). In Eq.(\ref{A}) $l$
is the AdS radius which is connected with the negative cosmological constant as $\Lambda =
-(d-1)(d-2) l^{-2}/2$.

The field strength reads
\begin{equation}\label{F}
F = \frac{1}{2} \bar{R}^{ab} J_{ab} + \frac{T^{a}}{l} P_{a}
\end{equation}
where $\bar{R}^{ab} = R^{ab} + l^{-2} e^{a} e^{b}$ with $R^{ab}=d\omega^{ab} +
\omega^{a}_{\hspace{1ex} c}\,\omega^{cb}$ the curvature two-form which contains the Riemann tensor
as $R^{ab} = \frac{1}{2} R^{ab}_{\hspace{1ex} cd} e^{c} e^{d}$. Finally $T^{a} = de^{a} +
\omega^{a}_{\hspace{1ex} c} e^{c}$ is the torsion two form.

For the anti de Sitter group $\langle F^{n}_{1} \delta A_{1} \rangle=0$ in
Eq.(\ref{variationoftrans}) implies in turns the two sets of equations
\begin{eqnarray}
\mathbf{G}_{a_{2n+1}} &=& \varepsilon_{a_{1}\ldots a_{2n+1}} \bar{R}^{a_{1}a_{2}}\ldots
\bar{R}^{a_{2n-3}a_{2n-2}}\bar{R}^{a_{2n-1}a_{2n}} = 0 \label{EE} \textrm{ and, }\\
\mathbf{T}_{a_{2n}a_{2n+1}} &=& \varepsilon_{a_{1}\ldots a_{2n+1}} \bar{R}^{a_{1}a_{2}}\ldots
\bar{R}^{a_{2n-3}a_{2n-2}}T^{a_{2n-1}} = 0 \label{Torsion},
\end{eqnarray}
where $\varepsilon_{a_{1}\ldots a_{2n+1}}$ is the $2n+1$ dimensional Levi-Civita antisymmetric
symbol.

Any locally AdS space ($\bar{R}^{ab}=0$ and $T^{a}=0$), in particular -global- AdS, solves
Eqs.(\ref{EE},\ref{Torsion}), however even these locally AdS solutions are far from being trivial
(see for instance \cite{Aminneborg:1996iz}). For instance the BTZ black hole \cite{Banados:1992wn}
belongs to this family of solutions.

\section{Conserved charges}

It is well known that the Noether formalism determines closed currents from the local symmetries
of a Lagrangian. In this case the action (\ref{Trans}) has two local symmetries to consider,
diffeomorphisms and gauge invariance. To review the Noether method see appendix
(\ref{NoetherMethod}).

The Noether current associated with gauge transformations can be directly derived using the
Noether prescription and considering the infinitesimal gauge transformation $A \mapsto A +
D(\lambda)$. That current reads \cite{ThesisPablomora}
\begin{equation}\label{Jgauge}
\hodge{{\bf J}_\lambda} = n(n+1)d \left( \int_{0}^{1} dt \langle (A_{1}-A_{0})F_{t}^{n-1} \lambda
\rangle \right).
\end{equation}
Analogously for diffeomorphism, $A \mapsto A + \lie_{\xi} A$, the current reads
\begin{equation}\label{Jdiff}
\hodge{{\bf J}_{\xi}}
= n(n+1) d\left( \int_{0}^{1}dt \langle (A_{1}-A_{0}) F_t^{n-1} I _{\xi }A _t\rangle \right).
\end{equation}

However it is not enough the existence of a current to have conserved charges. To have the very
concept of a conserved charge $\mathcal{M}$ must have, at least asymptotically, a timelike Killing
direction. In fact the charges are conserved along the timelike -asymptotical- Killing direction
of the manifold. Because of that in this work is assumed that the time like direction,
$\mathbb{R}$, satisfies this criterion. The generic form $\hodge{{\bf J}} = dQ$ in
Eqs.(\ref{Jgauge},\ref{Jdiff}) on the other hand splits the integral of the current into a charge
at $\partial \Sigma_{\infty}$ and another at $\partial \Sigma_{H}$.

\section{Boundary conditions}\label{Boundary Conditions}

\subsection{Spatial infinity}
As boundary condition in the asymptotical spatial region of $\mathcal{M}$, $\mathbb{R}\times
\partial \Sigma_{\infty}$, is imposed that the space be asymptotically locally AdS, which in turns implies that $A_{1}$ must satisfy
\begin{equation}\label{AsymptoticalCondition}
    A(x)_{1}|_{ x \rightarrow \partial\Sigma_{\infty} } \rightarrow \tilde{g}^{-1}d\tilde{g},
\end{equation}
where $\tilde{g}$ is an element of the AdS group.

Since $A_{0}$ already has the form $A_{0}=g^{-1} dg$ the condition (\ref{AsymptoticalCondition}),
might let $A_{1}$ and $A_{0}$ be globally unrelated, therefore the stronger condition
\begin{equation}\label{AsymptoticalConditionforreal}
A_{1}(x)|_{ x \rightarrow \partial\Sigma_{\infty} } \rightarrow A_{0}= g^{-1} dg
\end{equation}
is imposed. This condition states $A_{0}$ as a background field.

Since
\[
\Theta \propto (A_{1}-A_{0}),
\]
a direct consequence of this boundary condition is the vanishing of $\Theta$ at the asymptotical
spatial region $\mathbb{R}\times \partial \Sigma_{\infty}$ for arbitrary, but finite, variations
of the fields. In this way this boundary condition recovers in odd dimensions at least the basic
idea of the even dimensional ALAdS conditions \cite{Aros:1999id}.

\subsection{Horizon}

First one must recall that the horizon of a stationary/static black hole is a Killing horizon
defined as surface in $\mathcal{M}$ where the time like Killing vector, $\xi^{\mu}$, becomes light
like. As stated previously the horizon is to taken as an internal boundary.

By simple observation of the boundary term $\Theta$ in Eq.(\ref{Boundarytem}) one can note that
generically a boundary conditions at any surface can be $\delta A_{t}=0$. In this work this will
be exactly the boundary condition chosen at the horizon. Note that because $A_{0}$ is fixed
$\delta A_{t}=0 \Rightarrow \delta A_{1}=0$, and so this is actually a boundary condition for
$A_{1}$.

It is well known that the boundary conditions at the horizon are in direct connection with the
thermodynamical properties of the black hole \cite{Wald:1993nt}. In this particular case to fix
$A_{1}$ at the horizon determines the temperature of the black hole (See Appendix
\ref{GeoFisrtOrder}).

It is worth to mention that there is also a formal relation between $\omega^{ab}_{1}$ and the
extrinsic curvature. This leads to interpret the boundary term $B(A_{1},g^{-1}dg,F_{t})$ in
Eq.(\ref{TransRewritten}) as a generalization of the boundary term
\[
 \int_{\partial \mathcal{M}} (K-K_{0}) \sqrt{h}
\]
originally proposed in \cite{Hawking:1996fd}.

\section{Thermodynamics: Mass and Entropy}

Before to proceed it is necessary to recall the role the different boundaries play in the
discussion of charges. Mass, angular momentum or electric charge \cite{Aros:1999kt} are usually
identified as asymptotical values at -the spatial- infinity. On the other hand in the case of a
black hole geometry the charges at the horizon, the internal boundary,  are related with the
entropy of the black hole mainly \cite{Wald:1993nt}.

To proceed one particular black hole solution must be considered. Here it will be used the
solution obtained in \cite{Banados:1994ur,Aros:2000ij} ($T^{a}=0$), which can described by
vielbein
\begin{equation}
\label{vielbein} e^{0}=f(r) d\tau,\textrm{   } e^{1}= \frac{1}{f(r)}dr,\textrm{   } e^{m} = r
\tilde{e}^{m},
\end{equation}
and its associated torsion free connection
\begin{equation}
\label{spinconnection} \omega^{01}=\frac{1}{2}\frac{d}{dr}f(r)^{2} d\tau\,, \qquad \omega^{1m}=
-f(r) \tilde{e}^{m}\,, \qquad \omega^{mn} =  \tilde{\omega}^{mn},
\end{equation}
where
\[
f(r)^{2}= \gamma + r^2/l^2 -(\delta_{1 \gamma} + 2 M)^{1/n}.
\]

The transverse section is defined by the sub-vielbein $\tilde{e}^{m}$ and its associated torsion
free connection $\tilde{\omega}^{mn}$, with $m=2\ldots d-1$. The transverse section is a constant
curvature sub-manifold, i.e.,
\[
\tilde{R}^{mn} = \gamma \tilde{e}^{m} \tilde{e}^{n}
\]
with $\gamma =\pm 1, 0$.

For the geometry described by Eqs.(\ref{vielbein}), (\ref{spinconnection}) the generator of the
horizon, \textit{i.e.}, the Killing vector which defines the event horizon, simply reads
$\xi=\partial_{\tau}$. Note that since $e^{a}(\xi)=f(r)\delta^{a}_{0}$ the behavior near the
horizon can be analyzed in terms of the function $f(r)$, thus its larger zero, $f(r_{+})=0$,
determines the position of the horizon.

Now it rests to define the background field $A_{0}$. Recalling that $A_{0}$ must be a flat
connection one can define it analogously to Eqs.(\ref{vielbein},\ref{spinconnection}) by only
replacing $f_{0}(r)^{2} = \gamma + r^{2}/l^{2}$. It is direct to check that $F_{0}=0$, however in
order to $A_{0}$ be a proper background field the transverse section ( defined by
$\tilde{e}^{n},\tilde{\omega}^{mn} $) can not be arbitrary. Actually the locally constant
transverse sections that define genuine backgrounds are classified in \cite{Aros:2002rk}. Note
that the boundary condition $A_{1}\rightarrow A_{0}$ at the spatial infinity is manifest with this
choice.

\subsection{Variation of charges}

In general the Hamiltonian generator associated with a diffeomorphism $\eta$ can be written as
\[
G(\eta) = H(\eta) + \hat{G}(\eta)
\]
where $H(\eta)$ is a constraint, so it vanishes on shell, and $\hat{G}(\eta)$ a boundary term
which later one identifies with the value of the Hamiltonian charge associated with $\eta$.

To connect the Hamiltonian charges with the Noether charges one can follow the covariant phase
method \cite{Wald:1993nt}. The variation -on shell- of $\hat{G}$  in terms of the variation of
Noether charges formally reads
\begin{equation}\label{VariationofTheCharge}
\Xi(\hat{\delta},\delta_{\eta}) = \hat{\delta} \hat{G}|_{\partial \Sigma}=  \hat{\delta}
Q|_{\partial \Sigma} + \int_{\partial \Sigma} I_{\eta}\Theta(\hat{\delta}
A_{t},F_{t},A_{0},A_{1}),
\end{equation}
where $\hat{\delta}$ is transformation in the space of parameters of the solution and $\Xi$ is the
presymplectic form \cite{Wald:1993nt}. It is well established that in this case $\Xi$ vanishes
because $\delta_{\eta}=-\lie_{\eta}$ is a transformation of symmetry. The vanishing of $\Xi$ can
be understood as a conservation law between the spatial infinity and the horizon, i.e.,
\begin{equation}\label{XIconservation}
    \Xi |_{\partial \Sigma_{\infty}} = \Xi|_{\partial \Sigma_{H}}.
\end{equation}
This result is fundamental to obtain the first law of thermodynamics in this formalism
\cite{Wald:1993nt}.

\subsection{Charges at infinity}

A definition of mass can be obtained from the Hamiltonian charge associated with the time like
Killing vector $\xi =\partial_{\tau}$,\textit{ i.e.}, $\hat{G}(\partial_{\tau})|_{\partial
\Sigma_{\infty}}$. To obtain $\hat{G}(\partial_{\tau})|_{\partial \Sigma_{\infty}}$ one can use
the Noether current (\ref{Jdiff}) for the Killing vector $\xi =\partial_{\tau}$, whose value is
\begin{equation}\label{valueofcharges}
Q(\partial_{\tau})|_{\partial\Sigma_{\infty}} = \left(M + \frac{1}{2}\delta_{\gamma 1}\right),
\end{equation}
combined with Eq.(\ref{VariationofTheCharge}). Considering only the contribution from
$\partial\Sigma_{\infty}$ in Eq.(\ref{VariationofTheCharge}) one obtains
\begin{equation}\label{VariationofTheChargeatInfty}
\hat{\delta} \hat{G}(\partial_{\tau})|_{\partial\Sigma_{\infty}} =  \hat{\delta}
Q(\partial_{\tau})|_{\partial\Sigma_{\infty}}  + \int_{\partial \Sigma_{\infty}}
I_{\partial_{\tau}}\Theta(\hat{\delta} A_{t},F_{t},A_{0},A_{1}).
\end{equation}

However because of the boundary conditions in Eq.(\ref{AsymptoticalConditionforreal}) $\Theta$
vanishes at $\partial\mathcal{M}\mathbb{=R}\times
\partial\Sigma_{\infty}$ the last term in the r.h.s. of
Eq.(\ref{VariationofTheChargeatInfty}) vanishes yielding the direct relation
\[
\hat{\delta} \hat{G}(\partial_{\tau})|_{\partial\Sigma_{\infty}} =  \hat{\delta}
Q(\partial_{\tau})|_{\partial\Sigma_{\infty}},
\]
proving that Eq.(\ref{valueofcharges}) effectively represents the mass of the solution
(\ref{vielbein}). The rest of the Noether charges associated with the isometries of $\mathcal{M}$
vanish.

One can argue that in general the boundary conditions, in particular $A_{1}\rightarrow A_{0}$,
determine that Hamiltonian and Noether charges at infinity agree. In 2+1 dimensions is direct to
check that Eq.(\ref{Jdiff}) applied for the axial Killing vector $\partial_{\phi}$ in BTZ solution
gives the correct result. Unfortunately there is not known Chern Simons solutions with angular
momentum in higher dimensions to confirm this result.

\subsection{The horizon and the entropy}
As argued in Ref.{\cite{Wald:1993nt}} $\Xi_{H}$ yields the variation of the entropy of a black
solution. In this case the chosen $A_{0}$ determines the vanishing of the boundary term $\Theta$
not only asymptotically ($r\rightarrow \infty$) but for any value of $r>r_{+}$, \textit{i.e.},
\[
\Theta |_{\mathbb{R}\times \partial \Sigma_{r}}= \left.-n(n+1)\int^{1}_{0}dt \, \langle
(A_{1}-A_{0}) F_{t}^{n-1} \delta A_{t} \rangle\right|_{\mathbb{R}\times \partial \Sigma_{r}}
\equiv 0.
\]
where $\partial\Sigma_{r}$ stands for a surface at a finite fixed radius $r$.

This last result connects the Noether charge with the Entropy of this black hole since
$\Xi\equiv0$ \cite{Wald:1993nt} (see also \cite{Aros:2002ub}) implies that
\[
\Xi\equiv 0 \Rightarrow \underbrace{\delta M}_{ \Xi |_{\partial \Sigma_{\infty}}} = \underbrace{T
\delta S}_{\Xi|_{\partial \Sigma_{H}}}.
\]
After a straightforward calculation that yields
\begin{equation}\label{entropy}
\delta S = \frac{n}{2l} \left[\left(\gamma + \frac{r_{+}^{2}}{l^{2}}\right)^{n-1} \delta r_{+}
\right]\Sigma_{\gamma},
\end{equation}
reproducing the result found in \cite{Banados:1994ur} for these black holes.

\section{Gauge charges}

As for diffeomorphisms, for gauge transformations the Hamiltonian generator reads
\[
G(\lambda) = H(\lambda) + \hat{G}(\lambda),
\]
where $H(\lambda)$ is a constraint and $\hat{G}(\lambda)$ a boundary term representing value of
the Hamiltonian charge associated with $\lambda$.

As mentioned above the Noether procedure for the gauge invariance gives rise to the current
(\ref{Jgauge}). Throughout the covariant phase method one can readily confirm that the Noether
charge (from Eq.(\ref{Jgauge})) also agrees with Hamiltonian in this case, $\hat{G}_{\lambda}=
Q_{\lambda}$.

However one can also use the covariant phase method to compute the algebra of these charges.
Following \cite{Lee:1990nz} the formal expression for the gauge transformation of the gauge charge
reads
\begin{equation}\label{FomalBracket}
\delta_{\lambda_{1}} \hat{G}_{\lambda_{2}} = [\hat{G}_{\lambda_{1}},\hat{G}_{\lambda_{2}}] =
\hat{G}_{[\lambda_{1},\lambda_{2}]},
\end{equation}
showing that the algebra of charges reproduces the algebra of the underlying gauge group, in this
case AdS.

It is very interesting to compare this last result Eq.(\ref{FomalBracket}) with the one discussed
in Ref.\cite{Banados:1998ta} where, in 2+1 dimensions, a central charge arises because the
currents, in that case at the spatial infinity, are not gauge covariant. Therefore, one can argue
that the non existence of a central extension in Eq.(\ref{FomalBracket}) is due to the manifestly
gauge invariance of action (\ref{Trans}).

Another example of gravity in 2+1 dimensions in which a central extension arises is studied in
\cite{Brown:1986nw}. There the algebra of charges of diffeomorphisms at the spatial infinity also
has a central extension. However this result can be connected with the one for gauge charges. In
2+1 dimensions diffeomorphisms and gauge transformations are related in general, since on shell
$\lie_{\xi} A = D(I_{\xi} A)$. Furthermore, the standard EH action with a negative cosmological
constant, as mentioned before, is just the usual CS action with a different boundary term.

\section{Diffeomorphisms at the horizon}

In $d>3$ diffeomorphisms and gauge transformation are complete independent symmetries, thus the
study of the invariance of the boundary conditions under diffeomorphisms is interesting by itself.

Following the idea proposed in \cite{Carlip:1998wz} the diffeomorphisms that preserve the boundary
conditions at the horizon, \textit{i.e.}, the family of vector fields $\xi$ which satisfies
\begin{equation}\label{Boundarydiff}
-\left.n(n+1)\int^{1}_{0}dt \, \langle (A_{1}-A_{0}) F_{t}^{n-1} \lie_{\xi} A_{t}
\rangle\right|_{\mathbb{R}\times
\partial \Sigma_{H}} \equiv 0,
\end{equation}
can give rise to charges satisfying the Virasoro algebra with a central extension. In the original
prescription that central charge can be connected with the entropy of the black hole.

As discussed in \cite{Banados:1994qp} the analysis can be reduced just to the $\tau-r$ plane,
therefore the considered vectors have the generic form
\begin{equation}\label{xi}
    \xi = \xi^{\tau}(\tau,r) \partial_{\tau} + \xi^{r}(\tau,r)\partial_{r}.
\end{equation}

After straightforward computations one concludes that the condition (\ref{Boundarydiff}) leads to
an algebraical equation for the functions $\xi^{r}(\tau,r)$. If in additional one requires that
the Lie bracket of two vectors which satisfy the boundary conditions also satisfies this
condition, then the form of $\xi$'s vectors is obtained. After a Fourier decomposition and a
change of variable, the generic form of these vectors reads
\begin{equation}\label{xi-Al}
\xi_{m} = e^{im\tau} \kappa^{0}_{m} \left( \partial_{\tau} + X \partial_{X}\right) + O(X^{2}),
\end{equation}
where $\kappa^{0}_{m}$ is a constant, satisfying the constructive the relation
$\kappa^{0}_{m+n}=\kappa^{0}_{m}\kappa^{0}_{n}$, and $X=f_{1}(r)$ with $r>r_{+}$. These vectors
satisfy the algebra
\begin{equation}\label{Algebra}
    [\xi_{m},\xi_{n}] = i(m-n) \xi_{m+n}.
\end{equation}

The Noether current associated for diffeomorphisms in Eq.(\ref{Jdiff}) evaluated on the vector
Eq.(\ref{xi-Al}) gives rise at the horizon to the charges $\hat{G}_{\xi}= Q_{\xi}$, where
$Q_{\xi}$ arises from Eq.(\ref{Jdiff}). Using the covariant phase space formulation one can
compute algebra of these charges. The result of this procedure is
\begin{equation}\label{AlgebraInGeneral}
    [\hat{G}_{\xi},\hat{G}_{\eta}] = \hat{G}_{[\xi,\eta]} + K(\xi,\eta),
\end{equation}
where
\begin{equation}\label{central_in_gen}
K(\xi,\eta) = \int_{\partial \Sigma_{H}} n(n+1) \left( \int_{0}^{1}dt \kappa(\xi,\eta,t)\right)
\end{equation}
with
\[
\kappa(\xi,\eta,t) =  \langle\lie_{\eta} \left(( A_{1}-A_{0}) F_t^{n-1}\right) I _{\xi }A
_t\rangle - \langle I_{\eta}  \left(( A_{1}-A_{0}) F_t^{n-1}\right) \lie _{\xi }A _t\rangle.
\]

Therefore one can conclude that the algebra of charges indeed reproduces the algebra of
diffeomorphisms. Unfortunately one can confirm that for the solution Eq.(\ref{vielbein}) the
extension $K(\xi_{m},\xi_{n}) \sim O(X)$ while charges $\hat{G}_{\xi_{m}} \sim O(X^{0})$, which
implies that there is no central extension in the asymptotical limit at the horizon ($X\rightarrow
0$). This forbids to obtain an expression for the entropy in terms this central charge. A similar
result was obtained in \cite{Aros:2002ub}.

\section{Conclusions and remarks}

In this work the reinterpretation of the transgression form in Eq.(\ref{Trans}) as Chern Simons
action with a new set of boundary conditions has been discussed. The boundary conditions are not
determined by a particular group element but by a family of gauge elements related by gauge
transformations. In this way the general idea of a gauge theory is fully realized.

This boundary conditions allows to recover most of the usual structure as well as some standard
results. Furthermore, the boundary conditions allow that some of conclusions be even independent
of the particular background field $A_{0}$, as long as $A_{0}=g^{-1}dg$. For instance, although
different $A_{0}=g^{-1}dg$'s give rise to different values for the Noether charges as long as the
condition (\ref{AsymptoticalConditionforreal}) be satisfied the equivalence between the variations
of the Hamiltonian and Noether charges remains. Therefore the identification of mass or angular
momenta with the Noether charges associated with the global isometries is ensured independently of
the $A_{0}=g^{-1}dg$ chosen.

On the other hand, the fully gauge invariant action seems to be stripped off of all the rich
structure a Chern Simons theory has. In the standard Chern Simons theory the boundary degrees of
freedom, that arise because of CS action is a quasi invariant form, are fundamental in the
computation of the entropy \cite{Banados:1996tn}. Here, with a manifest gauge invariant theory,
the gauge transformations left no mark at the boundary which can contribute to the arise of new
degrees of freedom at the boundary.

One fundamental issue to study next is if the this fully gauge invariant action remains so beyond
tree level.

\appendix

\section{AdS group}\label{ads}

The AdS group can be defined as the set of transformations which leave invariant the quadratic
form $-x^{2}_{1}+ x^{2}_{2}+\ldots+x^2_{d}-x^{2}_{d+1}=-l^{2}$ in $d+1$ dimensions. Its generators
$J_{AB}$, with $A,B=1\ldots d+1$, satisfy the relation
\[
[J_{AB},J_{CD}] = - \delta_{AB}^{EG} \delta_{CD}^{FH} \eta_{EF} J_{GH}.
\]

Usually the trace of generator is normalized as
\[
\langle J_{A_{1} A_{2}} \ldots J_{A_{d} A_{d+1}} \rangle = \epsilon_{A_{1} A_{2} \ldots A_{d}
A_{d+1}},
\]
where $\epsilon_{A_{1} A_{2} \ldots A_{d}}$ is the $d+1$ dimensional Levi-Civita antisymmetric
symbol.

Finally in order to realize the group on a $d$ dimensional manifold the identification
\begin{equation}\label{indentification}
    J_{ab} = J_{ab} \textrm{ and } P_{a} = J_{a d+1}
\end{equation}
where $a=1 \ldots d$ can be used.

\section{Noether Method}\label{NoetherMethod}
An infinitesimal transformation of a field $\delta \phi(x)$ can be separated as (See
\cite{Ramond:1989yd})
\begin{equation}\label{SplitingTheTransf} \delta \phi(x)= \phi'(x') - \phi(x)= \phi'(x') -
\phi'(x) + \phi'(x) - \phi(x),
\end{equation}
where $\phi'(x) - \phi(x)= \delta_{0} \phi(x)$ is a local functional transformation, and
$\phi'(x') - \phi'(x) = {\cal L}_\xi \phi$  is the Lie derivative along $\xi$ (produced by the
diffeomorphism $x'=x + \xi(x)$). Thus $\delta\phi = \delta_{0}\phi + {\cal L}_\xi \phi$. Recalling
that $\lie_{\xi} = dI_{\xi}+ I_{\xi}d$ for a differential form the variation of a Lagrangian
$\mathbf{L}$ reads
\begin{equation}\label{LagrangianVariation}
  \delta {\mathbf{L}} = (E.M.)\delta_{0} \phi + d \Theta(\delta_{0}\phi,\phi) + d I_\xi
  {\mathbf{L}},
\end{equation}
where $(E.M.)$ stands for the \textit{equations of motion}.

A symmetry is defined as a change in the field configuration that leaves the equations of motion
invariant, which is satisfied if $\delta {\mathbf{L}} = d \alpha$. Therefore  the current
\begin{equation}\label{currentdensity}
 \hodge{{\bf J}_\xi} = \Theta(\delta_{0} \phi,\phi) + I_\xi{\bf L}-\alpha,
\end{equation}
evaluated on a solution satisfies $d(\hodge{{\bf J}_\xi})=0$. It is worth mentioning that for both
of the local symmetries of (\ref{Trans}) $\alpha=0$.

\section{Geometry in first order}\label{GeoFisrtOrder}

To fix $A_{1}$ at the horizon by Eq.(\ref{A}) in turns fixes $\omega^{ab}_{1}$. The connection
with the temperature is obtained by recalling that the horizon of a stationary/static black hole
is the surface where the generator of the horizon, the time like Killing vector $\xi^{\mu}$,
becomes light like. The vanishing of the norm of $\xi$ yields to the eigenvalue equation
\begin{equation}\label{Temperature}
I_{\xi} \omega^{a}_{\hspace{1ex} b} \xi^{b}|_{\mathbb{R}\times
\partial\Sigma_{H}} = \kappa \xi^{b},
\end{equation}
where $\kappa$ is the surface gravity and the temperature is given by $T=\kappa/2\pi$. This
relation is the first order version of
\[\xi^{\mu}\nabla_{\mu}(\xi^{\nu})|_{\mathbb{R}\times
\partial\Sigma_{H}} = \kappa \xi^{\nu} \]
obtained in \cite{Wald:1993nt}.

The fixing of $\omega_{1}^{ab}$ also the fixes the extrinsic curvature of $\mathbb{R}\times
\partial \Sigma_{H}$. To see that one can consider the $d$-dimensional orthonormal basis ($\eta,E_{I}$) (and the
corresponding vielbein $(e^{r}, e^{I})$). The vector $\eta$ defines a radial direction in which
the manifold can be foliated. The surfaces which are normal to $\eta$ have extrinsic curvatures
given by (See for instance \cite{yvonne})
\begin{equation}\label{extrinsical}
\kappa_{IJ} =  \eta_{a}(\nabla_{E_{I}} E_{J})^{a} =\eta_{r} \omega^{\,r}_{\hspace{1ex}IJ}.
\end{equation}
Equation (\ref{extrinsical}) is equivalent to the more standard definition
\begin{equation}\label{extrisical2}
\kappa_{\alpha \beta} = -\frac{1}{2}(\mathcal{L}_{\eta} g_{\alpha\beta})^{\parallel},
\end{equation}
where $g_{\alpha\beta}$ is the induced metric on a hypersurface and $^{\parallel}$ stands for the
parallel projection along that hypersurface.

{\bf Acknowledgments} I would like to thanks professors R. olea, R. Troncoso, J. Zanelli and in
particular P. Mora for enlightening discussions.  I would like to thank Abdus Salam International
Centre for Theoretical Physics (ICTP). This work was partially funded by grants FONDECYT 1040202
and DI 06-04. (UNAB).


\providecommand{\href}[2]{#2}\begingroup\raggedright\endgroup

\end{document}